\documentclass{article}

\usepackage[english]{babel}
\usepackage[letterpaper,top=2cm,bottom=2cm,left=3cm,right=3cm,marginparwidth=1.75cm]{geometry}

\usepackage{amsmath}
\usepackage{amssymb}
\usepackage{MnSymbol}
\usepackage{graphicx}
\usepackage[colorlinks=true, allcolors=blue]{hyperref}
\usepackage{subcaption}
\usepackage{siunitx}
\usepackage{float} 
\usepackage{natbib}
\usepackage{subcaption}

\usepackage{hyperref}
\usepackage{bm}
\usepackage{dcolumn}
\usepackage{tabstackengine}
\usepackage[dvipsnames]{xcolor}
\usepackage{ulem}
\usepackage{authblk}
\stackMath{}

\newcommand{\pd}[2]{\frac{\partial #1}{\partial #2}} 
\newcommand{\pdd}[2]{\frac{\partial^2 #1}{\partial #2^2}} 
\newcommand{\pdN}[3]{\frac{\partial^{#1} #2}{\partial #3^#1}} 
\newcommand{\intd}[1]{\textrm{d}#1}
\newcommand{\phiN}{\phi_{\textrm{n}}} 
\newcommand{\phiS}{\phi_{\textrm{s}}} 
\newcommand{\phiNbar}{\bar{\phi}_{\textrm{n}}} 
\newcommand{\phiNtil}{\hat{\phi}_{\textrm{n}}} 
\newcommand{\js}{{j}_{\textrm{s}}}

\newcommand{\LPatt}{L_{\textrm{patt}}}
\newcommand{\sigPrime}{\sigma'}
\newcommand{\sigStar}{\sigma^*}

\newcommand{\sub}[1]{_{\textrm{#1}}}

\newcommand{\Eqref}[1]{\mbox{equation\hspace{0.25em}\eqref{#1}}}
\newcommand{\Eqsref}[1]{\mbox{equations\hspace{0.25em}\eqref{#1}}}

\newcommand{\EQsref}[1]{\mbox{Equations\hspace{0.25em}\eqref{#1}}}
\newcommand{\figref}[1]{\mbox{figure\hspace{0.25em}\ref{#1}}}
\newcommand{\Figref}[1]{\mbox{Figure\hspace{0.25em}\ref{#1}}}

\newcommand{\secref}[1]{\mbox{section\hspace{0.25em}\ref{#1}}}
\newcommand{\secsref}[1]{\mbox{sections\hspace{0.25em}\ref{#1}}}

\newcommand{\appref}[1]{\mbox{appendix\hspace{0.25em}\ref{#1}}}

\begin{document}

\title{Dynamics of phase separation in non-local elastic networks}
\author{Oliver W. Paulin}
\author{Yicheng Qiang}
\author{David Zwicker}
\affil{Max Planck Institute for Dynamics and Self-Organization, Am Fa{\ss}berg 17, 37077 G\"{o}ttingen, Germany}

\date{\today}

\maketitle

\begin{abstract}
Phase separation of a liquid mixture embedded within an elastic network is relevant to a wide range of natural and industrial systems, including biomolecular condensates interacting with the cytoskeleton, structural colouring in bird feathers, and gas bubbles forming within soft sediments.
Recent experiments in synthetic polymer gels have demonstrated that when the size of phase-separated domains is comparable to the characteristic pore size of the network, a patterned phase with a well-defined length scale may emerge.
Theoretical works based on an equilibrium approach have attributed this pattern formation to non-local elastic effects arising from heterogeneity of the underlying network.
Here, we extend these ideas by developing a dynamic theory in which phase separation is coupled to non-local elasticity via the framework of large-deformation poroelasticity. 
We study our model via both linear stability analysis and numerical simulation, identifying the parameter space in which phase separation occurs, and investigating the impact of different elasticity models.
We find that although local elasticity can inhibit phase separation and affect domain count, it is unable to completely suppress coarsening.
In contrast, non-local elasticity arrests coarsening to form patterned domains with a well-defined length scale that decreases with increasing stiffness.
Our modelling framework thus paves the way for quantitative comparisons between simulations and experiments, for example by considering a strain-stiffening network rheology.
\end{abstract}

\tableofcontents

\section{Introduction}

The spontaneous phase separation of two immiscible fluids from a homogeneous mixture is a familiar everyday phenomenon.
Fluid--fluid phase separation of a mixture confined within the pore space of a deformable solid network is also widespread throughout natural and industrial settings.
Natural examples include sub-cellular condensates interacting with the cytoskeleton ~\citep{Wiegand2020, Lee2022, Zwicker2025}, and the formation of gas bubbles within soft geophysical materials such as seabed sediments and peatlands~\citep{Wheeler1988, Chen2015, Paulin2022}.
In materials science, phase separation in an elastic network can be utilised for the construction of novel micro-structured~\citep{FernandezRico2022, Fernandez-Rico2023} and stimulus-responsive~\citep{Celora2022, Celora2023} materials.
Finally, the texture and mechanical properties of many food products, such as chocolate and bread, depend sensitively on the size and distribution of small gas bubbles within the product~\citep{Pugh2023, Mathijssen2023}, which are in turn determined by the interactions of these bubbles with the surrounding solid material as they are formed.

Historical research into elastically mediated phase separation originated in the study of spinodal decomposition in hydrogel systems~\citep[\textit{e.g.}][]{Onuki1999}. 
Recently, significant interest in this field has been stimulated by experiments involving polymer gels saturated with an oil--oil mixture that phase separates into micron-scale droplets that exclude the elastic polymer network~\citep{Style2018, Rosowski2020, Rosowski2020b, Fernandez-Rico2023}.
It is now widely recognised that the additional energetic cost of network deformations induced by phase separation can inhibit the onset of phase separation, limit the size of phase-separated domains, and control domain morphology.
For gels in which there exists a gradient in material stiffness, phase separation will occur preferentially in softer regions of the gel.
This result can lead to an `elastic ripening' effect, in which droplets in softer regions grow at the expense of droplets in stiffer regions~\citep{Rosowski2020b,Vidal-Henriquez2020, Kothari2023}.
Further works have also studied the role of kinetic effects~\citep{Vidal-Henriquez2020} or a strain-stiffening network rheology~\citep{Kothari2020, Wei2020} in droplet size control, as well as investigating the impact of network damage~\citep{Vidal-Henriquez2021, Paulin2022} and wetting effects~\citep{Paulin2022, Ronceray2022, Liu2023}.

\begin{figure}
\centering
\includegraphics[width = 0.9\textwidth]{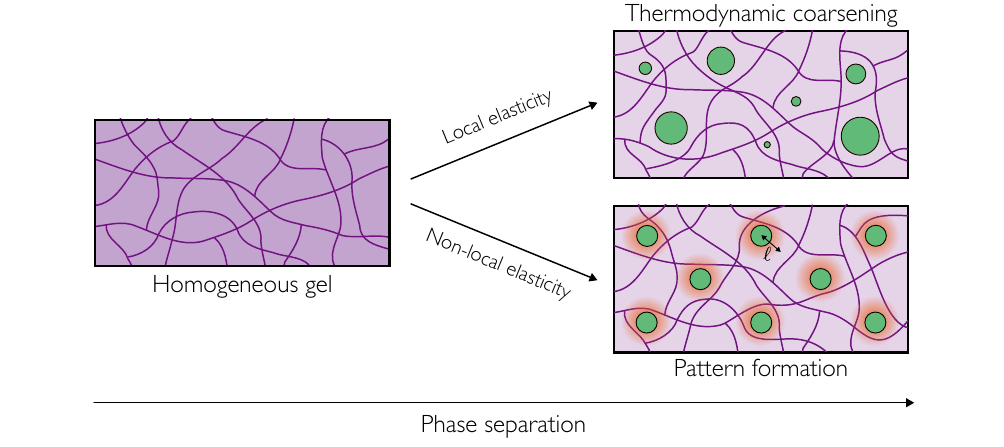}
\caption{Liquid-liquid phase separation within an elastic network (gel) can result in one of two distinct phenomenologies, depending on the type of elastic response. For a network whose response to deformation is well approximated by local elasticity, thermodynamic coarsening drives the system towards an equilibrium state consisting of a single phase-separated domain. In contrast, a non-local elastic response (over length $\ell$) inhibits coarsening, leading to selection of a finite pattern length scale.
}
\label{fig:Schematic}
\end{figure}

Recent experiments in the same system have also demonstrated the formation of persistent patterned phases that do not coarsen over time (\figref{fig:Schematic}), and that can form either droplet-like or bicontinuous morphologies~\citep{Fernandez-Rico2023}.
It has been hypothesised that these phases are equilibrium states that can form via a continuous phase transition from a homogeneous gel~\citep{Fernandez-Rico2023}.
Recent theoretical work has suggested that the observed size selection results from a competition between two different length scales: the elasto-capillary length; and a non-locality length arising from heterogeneities in the network mesh structure that manifest in a non-local elastic response~\citep{Qiang2024}.
Multi-dimensional extensions to this approach have studied the role of non-local elastic effects in controlling domain morphology~\citep{Mannattil2024, Oudich2025}.
The very recent work of \citet{Oudich2025} also explores dynamic simulations of this phenomenon, in a partially linearised elastic framework.
First steps towards quantitive comparisons between experiments and theory have been formulated by considering a strain-stiffening rheology for the elastic network~\citep{Fernandez-Rico2025}.

Here, we extend the equilibrium theory of \citet{Qiang2024} to study the dynamics of phase separation in a non-local elastic network, taking into account finite (large) deformations of the mesh.
Network deformation within swollen regions is typically extremely large, and so a theory that takes into account large deformations is needed.
Theories of phase separation in polymer gels that account for finite elastic deformations have previously been developed for studying the volume phase transition and spinodal decomposition in hydrogels~\citep{Onuki1999, Hong2008, Zhou2010, Hong2013, Drozdov2016, Bertrand2016, Hennessy2020}.
Our modelling approach represents an extension to these works that also incorporates non-local elastic effects. 
In this manuscript, we first layout our modelling framework (\secref{sec:Model}), deriving a one-dimensional (1D) model for simplicity. In \secref{sec:LSA}, we then conduct a linear stability analysis of our model, exploring the resulting dispersion relation and constructing phase diagrams. Next, we use our model to conduct numerical simulations of spontaneous phase separation, both for local (\secref{sec:Local}) and non-local (\secref{sec:NonLocal}) elastic responses. 
Finally, in \secref{sec:Conclusions} we discuss the implications of our work and connections to other similar studies.

\section{Model development} \label{sec:Model}

To construct our model, we consider a two-component mixture comprising an elastic network and a viscous solvent in 1D.
In order to describe the evolution of this mixture within the framework of large-deformation poroelasticity~\citep[\textit{e.g.},][]{MacMinn2016PRA, Hennessy2020}, we begin by defining kinematic relations between an Eulerian lab frame and a Lagrangian reference frame that is attached to the elastic network. 
Next, we prescribe a free energy functional that describes the energetic properties of the mixture, before constructing transport equations for the mixture via mass balance of each component.
Finally, we derive a constitutive law for the elastic stress in differential form that takes into account non-local elastic interactions.

\subsection{Kinematics}

We develop our model in the lab frame, with coordinates denoted by ${x}$. Coordinates in the reference frame are denoted by ${X}$.
Transformations between the two coordinate systems are governed by the deformation ${J}={\partial{x}}/{\partial{X}}$, which quantifies the ratio of volumes in the lab and reference coordinate systems.
Note that deformation is linked to the Eulerian displacement of the network ${u}({x})={x}-{X}\left({x}\right)$ via the relation  ${J}=\left(1-{\partial{u}}/{\partial{x}}\right)^{-1}$.
To characterise mixture composition, we define the Eulerian volume fraction of the network and solvent components as $\phiN({x})$ and $\phiS({x})$, respectively.
Assuming that both the network and solvent are locally incompressible, we can then write the no-void condition $\phiN+\phiS=1$, such that only one of $\phiN$ and $\phiS$ is required to fully determine the composition of the mixture. 
Finally, the volume fraction of a homogeneous, undeformed network ($J=1$) is given by the relaxed network fraction $\phiN(x)=\phiN^0$. 
Network fraction can thus be linked to deformation via ${J(x)=\phiN^0/\phiN(x)}$.

\subsection{Free energy}

To describe the energetic interactions of the network--solvent mixture, we follow \citet{Qiang2024} and assume that the total free energy $\Psi$ can be written as a sum of a local part and a non-local part,
\begin{equation}\label{eq:PsiTot}
\Psi = \int f_0\left(\phiN, \partial_x \phiN\right)\intd{x} + \int f\sub{nl}\left(J\right)\intd{X} \; ,
\end{equation}
where we define the local free energy denity $f_0$ in the lab frame, and the non-local energy density $f\sub{nl}$ in the reference frame.
We take $f_0$ to be a symmetric Flory--Huggins energy of the form
\begin{equation}
    f_0=\frac{k_\textrm{B}T}{\nu}\left[ \phiN\log\phiN + (1-\phiN)\log(1-\phiN)+\chi\phiN(1-\phiN) + \kappa \left(\pd{\phiN}{x}\right)^2 \right] \;.
\end{equation}
Here, $k_\textrm{B}$ is the Boltzmann constant, $T$ is temperature, $\nu$ is the molecular size, $\chi$ is the Flory interaction parameter, and $\kappa$ is an interfacial coefficient.
Note also  that the integral is taken over the volume element $\textrm{d}x$ defined in the lab frame.

For the elastic contribution to the energy, we assume that the network has a compressible Neo-Hookean rheology. For a local elastic model, this rheology is given by a hyperelastic strain-energy density
\begin{equation}
\mathcal{W}\sub{el} = \frac{G}{2}\left[J^2 - 1 - 2\ln J \right] + \frac{K}{2}\left(J-1\right)^2 \; ,
\end{equation}
where $G$ and $K$ are the respective shear and bulk moduli of the network.
We capture non-local elastic effects by taking the non-local energy to be a convolution of this energy with an isotropic kernel $g_\xi(X)$ in the reference frame.
The non-local free energy density is then
\begin{equation} \label{eq:PsiNL}
f\sub{nl} = \frac{G}{2} \left[ J \int J' g_\xi\left(X'-X\right)\intd{X}' -1-2\textrm{ln}\tilde{J} \right] + \frac{K}{2} (J-1) \int (J' -1) g_\xi\left(X'-X\right)\intd{X'}  \;,
\end{equation}
where $\tilde{J} =  \int J' g_\xi\left(X'-X\right)\intd{X'}$, $J'$ is the value of $J$ evaluated at $X'$, and $\textrm{d}X $ is the volume element in the reference frame. We use this expression to find a non-local constitutive law for the elastic stress in \secref{sec:NonLocalLaw} below.

\subsection{Transport equations}

We now derive transport equations describing the evolution of mixture composition. To begin, we consider conservation of mass of the network, writing the continuity equation in the Eulerian frame as
\begin{equation} \label{eq:NetworkConservation}
    \pd{\phiN}{t}+\pd{}{x}(\phiN {v}\sub{n})=0 \; ,
\end{equation}
where ${v}\sub{n}=J\partial{u}/\partial t$ is the Eulerian velocity of the network.
Similarly, conservation of mass for the solvent implies that
\begin{equation} \label{eq:SolventConservation}
    \pd{\phiS}{t}+\pd{}{x}(\phiS {v}\sub{s})=0 \; ,
\end{equation}
where ${v}\sub{s}$ is the Eulerian solvent velocity.
Using $\phiN+\phiS=1$, and summing \Eqsref{eq:NetworkConservation} and \eqref{eq:SolventConservation} then gives $\partial q/\partial x = 0$, 
where ${q}=\phiN{v}\sub{n}+\phiS{v}\sub{s}$ is the total mixture flux. In 1D, $q$ is constant, and is determined by the boundary conditions. 
Defining the flux of solvent relative to the network as $\js=\phiS({v}\sub{s}-{v}\sub{n})$, we can rewrite \Eqref{eq:NetworkConservation} in terms of ${q}$ and $\js$ as
\begin{equation} \label{eq:NetworkEvolution}
    \pd{\phiN}{t}+\pd{}{x}\left[\phiN({q}-\js)\right]=0 \; .
\end{equation}
Given an appropriate expression for the solvent flux $\js$, \Eqref{eq:NetworkEvolution} then provides a complete description of the dynamics of the mixture.
Note that by focussing on a system with periodic boundary conditions, we can set $q$ to zero without loss of generality.

To find an expression for $\js$, we assume that fluxes are driven by gradients in both chemical potential $\mu$ and elastic stress ${\sigma}$~\citep{Onuki1999},
\begin{equation} \label{eq:jDef}
    \js=-\Lambda \left( \pd{\mu}{x}+\frac{1}{\phiN}\pd{{\sigma}}{x}\right) \; ,
\end{equation}
where $\Lambda(\phiN)$ is the solvent mobility.
The exchange chemical potential $\mu$ is calculated as the variational derivative of the free energy with respect to composition such that $\mu=-\nu{\delta \Psi}/{\delta \phiN}$. For the free energy defined in \Eqref{eq:PsiTot}, this gives
\begin{equation} \label{eq:chempot}
\mu = {k_\textrm{B}T}\left[ \log(1-\phiN) -\log{\phiN} - \chi(1-2\phiN) + \kappa\pdd{\phiN}{x}\right] \;.
\end{equation}
Similarly, an expression for the elastic stress ${\sigma}$ in terms of deformation $J$ is found by taking the derivative of energy with respect to deformation, as discussed in detail below.

\subsection{Non-local stress law} \label{sec:NonLocalLaw}

To quantify the elastic stress in the Eulerian frame, we work with the Cauchy stress tensor, defined as
\begin{equation} \label{eq:sigDef}
{\sigma} = \frac{\delta \Psi}{\delta J} \; .
\end{equation}
For the non-local elastic energy defined in \Eqref{eq:PsiNL}, we thus find that
\begin{equation}
\sigma = G \left[ \int J' g_{\xi}\intd{X'} - \frac{1}{\tilde{J}}\right] + K\int(J'-1)g_{\xi}\intd{X'} \; ,
\end{equation}
where we recall that $\tilde{J} =  \int J' g_\xi\intd{X}$.
For a given convolution kernel, it may be possible to write an equation for this non-local stress in differential form. 
To do this, it is convenient to first split the elastic stress into two parts, such that $\sigma = \sigPrime - \sigStar$, where $\sigPrime = \int\left[ GJ' + K(J'-1) \right]g_\xi\intd{X'}$ and $\sigStar = G/\tilde{J}$.
We then choose the convolution kernel $g_\xi=\exp({-|{X}|/\xi})/2\xi$, where $\xi$ is the non-locality scale.
In this case, we find that
\begin{equation} \label{eq:sigPrimeEq}
\left(1- \xi^2\pdd{}{X}\right)\sigPrime = GJ + K(J-1) \;
\end{equation}
and
\begin{equation} \label{eq:sigStarEq}
\left(1- \xi^2\pdd{}{X}\right)\frac{1}{\sigStar} = \frac{J}{G} \;,
\end{equation}
where $\partial/\partial X=J\partial / \partial x$ is the spatial derivative with respect to coordinates in the reference state, and we recall that $J$ is linked to  network volume fraction via the kinematic relation  ${J = \phiN^0/\phiN}$.
Note that in the limit of zero non-locality length $\xi=0$, we recover the local Neo-Hookean stress $\sigma = G(J - 1/J) + K(J-1)$.
A detailed derivation of these equations is provided in \appref{App:Stress}.

\subsection{Model summary and non-dimensionalisation}

\EQsref{eq:NetworkEvolution}--\eqref{eq:chempot}, along with \Eqsref{eq:sigPrimeEq} and \eqref{eq:sigStarEq}, provide a complete set of coupled partial differential equations that describe the evolution of the network--solvent mixture.
Non-dimensionalising our model by length scale $\sqrt{\kappa}$, time scale $\kappa\nu/ \Lambda k\sub{B}T$, and stress scale $k\sub{B}T/\nu$ motivates introducing the dimensionless variables $\mathcal{G} = G\nu/k\sub{B}T$, $\mathcal{K} = K\nu/k\sub{B}T$, and $\ell^2=\xi^2/\kappa$. Here, $\mathcal{G}$ and $\mathcal{K}$ are the rescaled shear and bulk moduli of the elastic network, respectively, and $\ell$ quantifies the size of the non-locality length relative to the interfacial width $\sqrt{\kappa}$.
Choosing a solvent mobility of the form $\Lambda(\phiN) = \Lambda (1-\phiN)\phiN$, we now analyse our model via both linear stability analysis (\secref{sec:LSA}) and numerical simulations (\secsref{sec:Local} and \ref{sec:NonLocal}).
To carry out numerical simulations, we discretise in space on a uniform grid, and approximate spatial derivatives with a centralised finite difference scheme. We then solve for $\phiN$, $\sigPrime$ and $\sigStar$ at each time step with MATLAB's in-built ode solver \texttt{ode15s}.

\section{Linear stability analysis} \label{sec:LSA}

To gain insight into the parameters that control the onset of phase separation, we now conduct a linear stability analysis of our model. 
We linearise our system about a base state corresponding to a uniformly swollen network with volume fraction $\phiNbar$. 
The corresponding base stress state is then $\bar{\sigma}=\mathcal{G}\left(\bar{J}-1/\bar{J}\right) + \mathcal{K}\left( \bar{J} -1 \right)$, where $\bar{J} = \phiN^0/\phiNbar$.
Setting $\phiN=\phiNbar \left(1 + \delta \phiNtil \right)$ and $\sigma=\bar{\sigma} \left(1 + \delta \hat{\sigma}\right)$, where $\delta\ll1$, we arrive at the following linearised equation for the evolution of perturbations in network volume fraction
\begin{equation}
\pd{\phiNtil}{t}=- (1-\phiNbar) \left[\phiNbar\left(\Delta f\pdd{}{x} + \pdN{4}{}{x} \right) \phiNtil + \pdd{\hat{\sigma}}{x}  \right]\;,
\end{equation}
where $\Delta f=2\chi - \phiNbar^{-1} - (1-\phiNbar)^{-1}$ is the linearised driving force of phase separation. Similarly, we find the linearised stress equation
\begin{equation}
\left[1 - \left(\frac{\ell\phiN^0}{\phiNbar} \right)^2  \pdd{}{x} \right]\hat{\sigma} = \left[ \mathcal{G} \left( \frac{\phiN^0}{\phiNbar} + \frac{\phiNbar}{\phiN^0}\right) + \mathcal{K} \frac{\phiN^0}{\phiNbar}\right] \phiNtil \; .
\end{equation}
Substituting the ansatz $\phiNtil,~\hat{\sigma} \propto \exp\left(st+ikx\right)$, where $s$ is the growth rate of perturbations with wavenumber $k$, we arrive at the dispersion relation
\begin{equation} \label{eq:Dispersion}
s =  \Delta f k^2 - k^4 - \frac{\beta k^2}{1+\alpha^2k^2} \; ,
\end{equation}
with $\beta =\mathcal{G} \left( {\phiN^0}/{\phiNbar^2} + {1}/{\phiN^0}\right) + \mathcal{K} {\phiN^0}/{\phiNbar^2}$ a rescaled stiffness, and $\alpha=\ell\phiN^0/\phiNbar$ a rescaled non-locality scale.
If $s(k)>0$, a uniformly swollen network will be unstable to perturbations of wavenumber $k$. As such, the homogeneous base state is only stable if $s<0$ for all $k$.
From \Eqref{eq:Dispersion}, we see that the interfacial ($k^4$) term is always stabilising, and dominates at large $k$.
The elastic contribution ($\beta$ term) is also stabilising for all wavenumbers, and thus acts to inhibit the onset of phase separation.
In contrast, the thermodynamic driving force may be either stabilising or destabilising depending on the sign of $\Delta f$. 

\begin{figure}
\centering
\includegraphics{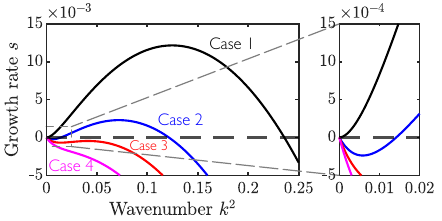}
\caption{Dispersion relation showing growth rate $s$ as a function of squared wavenumber $k^2$. We identify four distinct functional forms of $s(k)$, labelled cases 1 to 4. The inset highlights the behaviour of $s(k)$ at small $k$.
}
\label{fig:Dispersion}
\end{figure}

Depending on the relative magnitudes of $\Delta f$, $\beta$, and $\alpha$, we identify four distinct functional forms for $s(k)$, labelled as cases 1 to 4 in \figref{fig:Dispersion}.
Case 1 occurs when $s'(k=0)= \Delta f - \beta>0$, and corresponds to the classical spinodal instability.
In case 2, $s'(k=0)<0$, but $s>0$ for some compact band of wavenumbers $k\in[k_1,k_2]$. In this case, we expect instabilities to arise at a well-defined finite length scale. We refer to this possibility as the `patterned instability'.
For both case 3 and 4, $s<0$ for all $k$, and so the homogeneous base state is stable. 
However, $s(k)$ may still have a local maximum away from $s=0$ (case 3), thus providing a length scale that is less stable than neighbouring modes.
In a noisy system, this mode may be able to induce an instability and nucleate phase-separated domains.
In \secsref{sec:LocalLSA} and \ref{sec:NonLocalLSA} below, we use the instability criteria derived here to reveal how the onset of these instabilities depends on different model parameters, for both local (\secref{sec:LocalLSA}) and non-local (\secref{sec:NonLocalLSA}) elasticity laws.

\section{Phase separation with local elasticity} \label{sec:Local}

Before investigating the predictions of our full model, we first focus on the limiting case of local elasticity ($\ell=0$), to compare to the results of previous studies. 
Previous results have shown that elasticity can suppress phase separation entirely~\citep{Wei2020, Kothari2020, Hennessy2020, Paulin2022}, limit the final size of phase-separated domains~\citep{Vidal-Henriquez2020, Vidal-Henriquez2021, Wei2020}, and decrease the rate of post-separation coarsening~\citep{Onuki1999}.

\subsection{Elasticity inhibits the onset of phase separation} \label{sec:LocalLSA}

To begin, we use the linear stability analysis presented in \secref{sec:LSA} to identify regions of the parameter space in which we expect a homogeneous mixture with network fraction $\phiNbar$ to phase separate.
For a local elasticity law, the dispersion relation (equation \ref{eq:Dispersion}) simplifies to $s(k) = (\Delta f -\beta) k^2 - k^4$. As such, only case 1 (classical spinodal instability) and case 4 (stable homogeneous state) of those identified in \secref{sec:LSA} are possible.

\begin{figure}
\centering
\includegraphics{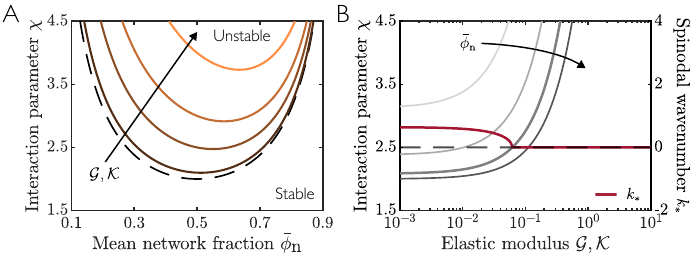}
\caption{Spinodal curves for a local elastic model ($\ell=0$) with $\phiN^0=1$ calculated via linear stability analysis. (A) Dark to light curves show spinodals for increasing elastic moduli $\mathcal{G}, \mathcal{K}$. The dashed black curve shows the limiting case $\mathcal{G}=\mathcal{K}=0$.
(B) Light to dark grey curves show spinodals for increasing mean network fraction $\phiNbar$. The thick red line shows the most unstable wavenumber $k_*$ as a function of $\mathcal{G},\mathcal{K}$ for fixed $\chi = 2.5$, $\phiNbar=0.4$. The spinodal curve for $\phiNbar=0.4$ is highlighted by a thicker line width.
}
\label{fig:LocalLSA}
\end{figure}

For local elasticity, the spinodal curve is then defined by $\Delta f = \beta$. 
When $\mathcal{G}, \mathcal{K}=0$ ($\beta = 0$; no elasticity), the standard Flory--Huggins spinodal $2\chi = \phiNbar^{-1} + (1-\phiNbar)^{-1}$ is recovered.
Inside the spinodal curve, a homogenous mixture will be unstable to small perturbations and phase separate into domains of high and low network fraction.
As the impact of elasticity increases (increasing $\mathcal{G}, \mathcal{K}$), the size of the spinodal region decreases (\figref{fig:LocalLSA}A).
Consistent with previous results~\citep{Wei2020, Vidal-Henriquez2020, Hennessy2020, Paulin2022}, the additional energetic cost associated with deforming the elastic network inhibits the onset of phase separation. 
Note that the asymmetry of the spinodal curve about $\phiNbar=0.5$ results from taking the initial homogeneous network to be swollen relative to the reference state ($\phiNbar<\phiN^0$), therefore providing a bias away from small network fractions.
\Figref{fig:LocalLSA}B shows the spinodal curve as a function of elastic moduli $\mathcal{G}, \mathcal{K}$ and interaction parameter $\chi$, highlighting that phase separation is strongly inhibited at large stiffness.

\subsection{A stiffer network leads to fewer droplets} \label{sec:LocalDynamics}

\begin{figure}
\centering
\includegraphics[]{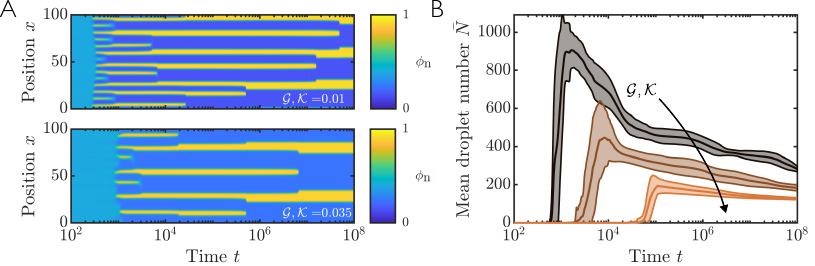}
\caption{Numerical simulations of phase separation within a local elastic network. 
(A) Example simulations showing network fraction $\phiN(x, t)$ for a soft ($\mathcal{G}, \mathcal{K} = 0.01$; top) and stiff ($\mathcal{G}, \mathcal{K} = 0.035$; bottom) elastic network. Note that the results shown here are a representative snapshot of the full simulation domain $x\in[0, 10^4]$.
(B) Mean number of droplets as a function of time for $\mathcal{G,K}=0, 0.01, 0.02$ (dark to light colours). Thick solid lines show mean results from $100$ simulations with different initial conditions, and shaded regions show one standard deviation either side of this mean.
Other parameter values are $\phiNbar=0.4$ and $\chi=2.5$.
}
\label{fig:LocalCoarsening}
\end{figure}

To study the time evolution of the mixture predicted by our model, we now perform numerical simulations of \Eqsref{eq:NetworkConservation}--\eqref{eq:sigStarEq} with $\ell=0$. 
\Figref{fig:LocalCoarsening}A shows example simulation results for both a soft (top) and stiff (bottom) network.
In both cases, a homogeneous mixture spontaneously phase separates into distinct solvent-rich (droplet; $\phiN\sim0$) and network-rich ($\phiN\sim1$) domains, via the classical spinodal instability. 
Over time, these domains then coarsen, with the number of droplet domains decreasing (\Figref{fig:LocalCoarsening}B).
Although coarsening is exponentially slow in 1D, we expect the eventual steady state to consist of a single droplet to minimise total interfacial area.

We find that for a stiff network, mean droplet size is generally slightly larger than for a soft network.
However, this size difference is counteracted by a vastly smaller total number of droplets in the stiff case, reconciling the expectation that the total volume of solvent droplet in the system is smaller for a stiffer material.
\Figref{fig:LocalCoarsening}B shows the mean droplet number calculated from many simulations for different elastic moduli $\mathcal{G}, \mathcal{K}$. We find that the coarsening rate appears relatively insensitive to stiffness, but that the onset of phase separation is slightly delayed for stiffer networks.
The initial droplet spacing is set by the spinodal wavenumber $k_*$, which can be predicted from the most unstable wavenumber (largest $s$) of the linear stability analysis. \Figref{fig:LocalLSA}B shows $k_*$ as a function of stiffness for fixed $\phiNbar$ and $\chi$.
We see that $k_*$ decreases with increasing $\mathcal{G}, \mathcal{K}$, implying that droplets will be more spaced out, hence leading to fewer droplets, in line with the observations from both our numerical simulations and previous studies.

\section{Phase separation with non-local elasticity} \label{sec:NonLocal}

Next, we explore the additional impact of non-local elasticity on phase separation by studying our full model, via both linear stability analysis and numerical simulations. These results extend the work of \citet{Qiang2024} by considering the full temporal evolution of the network--solvent mixture.

\subsection{Phase plane reveals different modes of instability} \label{sec:NonLocalLSA}

\begin{figure}
\centering
\includegraphics[width = \columnwidth]{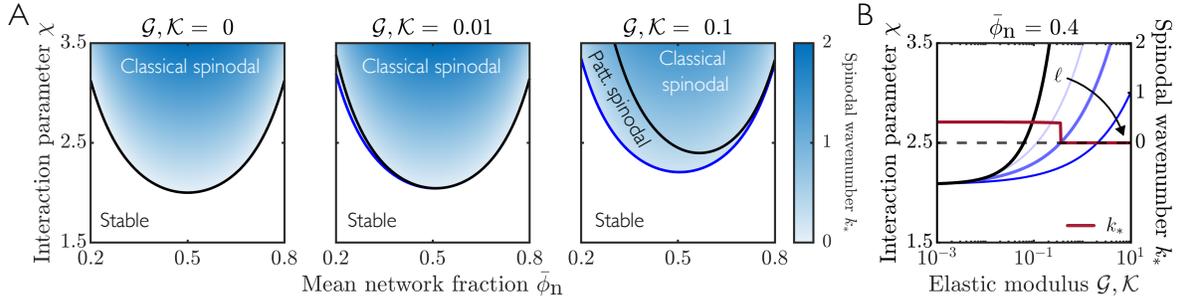}
\caption{Phase planes resulting from a linear stability analysis of the full non-local model. Black curves indicate the classical `macroscopic' spinodal, and blue curves indicate the `patterned' spinodal.
(A) Spinodal curves for different elastic moduli $\mathcal{G}, \mathcal{K}$, with $\ell=2$. Blue shading identifies the most unstable wavenumber $k_*$ at each point within the spinodal.
(B) Light to dark blue curves show patterned spinodals corresponding to non-locality length $\ell = 1$, $2$, $5$ for fixed $\phiNbar=0.4$. The macroscopic spinodal (black curve) is the same for all values of $\ell$.
The thick red curve shows the most unstable wavenumber $k_*$ as a function of $\mathcal{G}, \mathcal{K}$ for fixed $\chi = 2.5$, $\phiNbar=0.4$, $\ell=2$. The spinodal curve for $\ell=2$ is highlighted by a thicker line width for clarity.
}
\label{fig:NonLocalLSA}
\end{figure}

Using our linear stability analysis, we now identify the stable and unstable regions of the parameter space  for the non-local model (\figref{fig:NonLocalLSA}). As for a local elastic response, we see that increasing network stiffness increases the stability of the mixture, and hence inhibits the onset of phase separation.
In contrast to the local elastic case, however, the inclusion of non-local elasticity allows for two distinct modes of instability, as identified in \secref{sec:LSA}.
When $s'(k=0)>0$ (case 1), all sufficiently long wavelengths are unstable to perturbations, resulting in classical spinodal decomposition. The boundary of the region in which this instability occurs is marked by the black curves in \figref{fig:NonLocalLSA}.
When $s'(k=0)<0$ (case 2), a discrete band of unstable wavenumbers is now possible, with large wavelength modes now stable.
In addition to the classical macroscopic spinodal identified in \secref{sec:Local}, we can thus also identify a `patterned' spinodal region (\figref{fig:NonLocalLSA}; blue line) that corresponds to this case.
A homogeneous state is then unstable to perturbations if it is inside either the classical or patterned spinodals.

As stiffness $\mathcal{G}, \mathcal{K}$ increases, the size of the macroscopic spinodal region decreases significantly, but the size of the `patterned region' between the macroscopic (black line) and patterned (blue line) spinodals increases.
Non-local elastic effects thus increase the range of parameters for which the homogeneous state is unstable, promoting phase separation relative to local elasticity.
Increasing the non-locality length $\ell$ amplifies this effect (\figref{fig:NonLocalLSA}B) by further expanding the boundary of the patterned spinodal.

\Figref{fig:NonLocalLSA}B also shows the most unstable wavelength $k_*$ predicted by the linear stability analysis as a function of stiffness, for fixed $\chi$ and $\phiNbar$. A notable feature of this result is that $k_*$ remains finite at the boundary of the patterned region.
This feature is generically true for all points along the patterned spinodal. In contrast, $k_*=0$ where the boundary of the unstable region is defined by the macroscopic spinodal (both for local and non-local elasticity).
Finite $k_*$ at the spinodal implies that a continuous quench across this boundary will preferentially induce a finite wavelength instability.
This result raises the possibility of a continuous phase transition across the patterned spinodal, as predicted previously via an equilibrium approach~\citep{Qiang2024}.

\subsection{Non-local elasticity leads to arrested coarsening}

\begin{figure}
\centering
\includegraphics{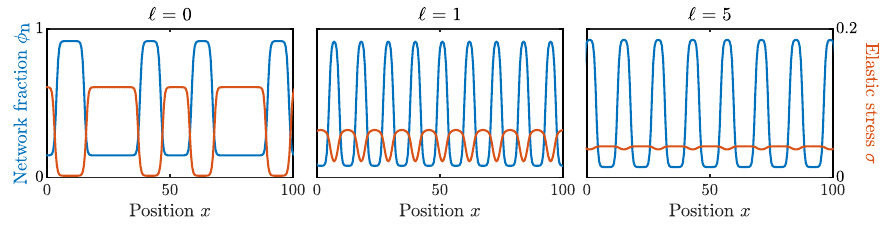}
\caption{Spatial distribution of network fraction $\phiN$ (blue curves) and elastic stress $\sigma$ (orange curves) at the final time point of numerical simulations, for different non-locality lengths $\ell=0$, $1$, $5$. Simulation results are shown at time $t\sub{end}=10^8$ and are representative snapshots of the full simulation domain $x\in[0, 2\times10^4]$. Other parameter values are $\phiNbar=0.4$, $\mathcal{G}=\mathcal{K}=0.01$ and $\chi = 3$.
}
\label{fig:NonLocalStress}
\end{figure}

To investigate the long term dynamics of a phase separating mixture, we now turn to numerical simulations of our model.
We simulate \Eqsref{eq:NetworkConservation}--\eqref{eq:sigStarEq} up until a fixed end time $t\sub{end}$, starting from an initial condition corresponding to a homogenous mixture superimposed with small random perturbations.
\Figref{fig:NonLocalStress} shows the spatial distribution of network fraction $\phiN$ and elastic stress $\sigma$ at the end time of our simulations for various values of the non-locality length~$\ell$. Note that each plot shows only a small snapshot of the full simulation domain.

\begin{figure}
\centering
\includegraphics{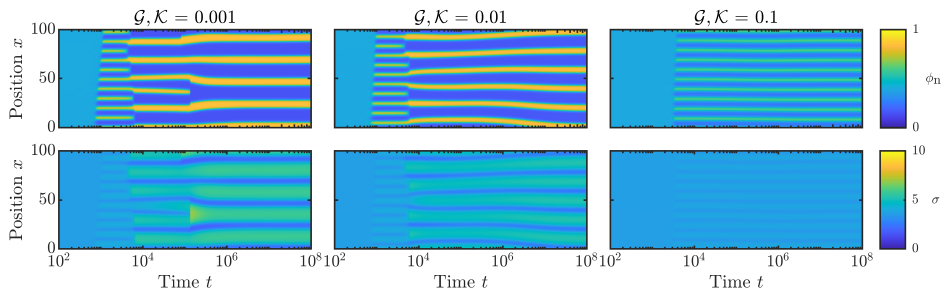}
\caption{Kymographs showing network fraction $\phiN$ (top) and elastic stress $\sigma$ (bottom) as functions of space and time for numerical simulations with different elastic moduli $\mathcal{G}, \mathcal{K}$. These results show a representative snapshot of the full simulation domain $x\in[0, 2\times10^4]$, and parameter values are the same as for \figref{fig:NonLocalStress}, with $\ell=2$.
}
\label{fig:NonLocalDynamics}
\end{figure}

For $\ell=0$ (local elastic limit), we find phase-separated domains of alternating high and low $\phiN$ with irregular size and spacing.
If simulations were run for a longer time, we would expect these domains to continue to coarsen to the thermodynamic equilibrium of a single domain. Since coarsening is exponentially slow in 1D, we instead choose to end our simulations at $t\sub{end}$ before this final state is reached, and highlight the fact that no regular patterned structure is present in this case.
We also note that $\sigma$ correlates exactly with $\phiN$, with regions of low network fraction corresponding to high tensile stress. In regions of large $\phiN$, the elastic stress is relatively much smaller, but still in tension since the initial condition corresponds to a highly swollen network relative to the relaxed network fraction $\phiN^0=1$.
For $\ell=1$ and $\ell=5$, however, we find a regular periodic, or patterned, state at $t=t\sub{end}$, which does not evolve further with time.
Here, the non-local elastic contribution spreads the elastic stress over a finite spatial width, leading to much smaller variations in $\sigma$ than those observed for $\phiN$.
As $\ell$ increases further, the stress becomes increasingly uniform, and so the pattern length scale increases.
In this case, our simulations have reached a steady state by the end of the simulation time, and do not coarsen further.

\Figref{fig:NonLocalDynamics} shows $\phiN$ (top) and $\sigma$ (bottom) as functions of space and time for different elastic stiffness at fixed non-locality length $\ell=2$. 
As before, the results presented here represent a small snapshot of the full simulation domain.
At early times, we see that phase separation initiates at the spinodal length scale, forming many small droplets of low network fraction.
Initial phase separation is then followed by partial coarsening, with large droplets growing at the expense of small droplets.
Eventually, however, coarsening is completely arrested and the system reaches a steady patterned state.
This patterned state also persists for much longer simulation times than those presented here.
The same phenomenology is observed for all value of elastic stiffness $\mathcal{G}, \mathcal{K}$, but a larger stiffness results in a smaller pattern length scale.
We denote the pattern length scale that emerges from these simulations by $\LPatt$.
We note that the value of $\LPatt$ is robust to variations in the size of the simulation domain, indicating that this is an energetically selected length scale, rather than being set by the finite size of our simulations.

\begin{figure}
\centering
\includegraphics{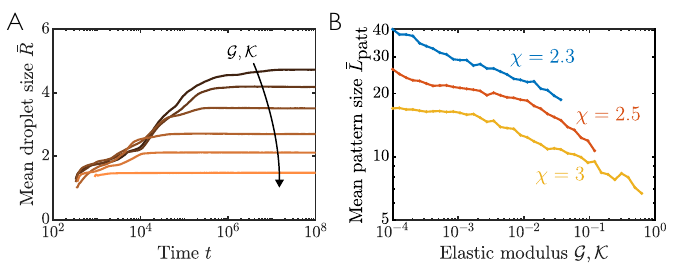}
\caption{(A) Mean droplet size $\bar{R}$ from $100$ simulations, for different elastic moduli varying between $\mathcal{G}, \mathcal{K} = 10^{-4}$ and $0.3$ (dark to light curves). Other parameter values are $\phiNbar=0.4$, $\chi=3$ and $\ell=2$.
(B) Mean pattern length ${\bar{L}\sub{patt}}$ as a function of elastic modulus $\mathcal{G}, \mathcal{K}$ for different values of the interaction parameter $\chi$. Mean values are taken from $100$ simulations, and parameter values are the same as for panel (A). Note that this plot is displayed on a double logarithmic scale.
}
\label{fig:NonLocalCoarsening}
\end{figure}

To analyse the dependence of the emergent pattern length scale on different model parameters, we now repeat our simulations for $100$ sets of different realisations of the random perturbations applied to the initial condition.
\Figref{fig:NonLocalCoarsening}A shows the mean size of droplet domains $\bar{R}$ as a function of time.
As in \figref{fig:NonLocalDynamics}, we see that initial coarsening is eventually arrested, with the mean droplet size then remaining constant.
In contrast to the spinodal length scale (as discussed in \secref{sec:LocalDynamics}), we find that a stiff network (large $\mathcal{G}, \mathcal{K}$) leads to a smaller pattern length scale than a soft network (small $\mathcal{G}, \mathcal{K}$), and thus a smaller mean droplet size and larger droplet count.
We also observe that a larger stiffness leads to earlier arrest of coarsening.
For extremely soft networks, a steady patterned state is not reached within the timeframe of our simulations.
For sufficiently large stiffness such that the system is initialised in the patterned region identified in \secref{sec:NonLocalLSA}, we find that the mixture immediately phase separates into the patterned state and thus does not coarsen at all.
\Figref{fig:NonLocalCoarsening}B highlights the dependence of $\LPatt$ on stiffness explicitly.
Although our results show $\LPatt$ decreasing with stiffness, we note that this dependence is weaker than the square root scaling predicted by previous equilibrium theories~\citep{Qiang2024, Mannattil2024}.
We hypothesise that this discrepancy could result from simulations becoming trapped in a non-equilibrium patterned state, due to the finite system size and periodic boundary conditions.
We also repeat our simulations for different values of the interaction parameter $\chi$. For all values of $\chi$ we see the same qualitative behaviour, but with smaller values of $\chi$ resulting in a larger pattern length scale (\figref{fig:NonLocalCoarsening}B).

\section{Conclusions} \label{sec:Conclusions}

Recent experiments with synthetic polymer gels, as well as potential applications to biomolecular condensates, have fuelled significant interest in elastically mediated fluid--fluid phase separation.
The presence of a confining elastic network that is permeated by a fluid mixture induces an additional energetic cost to the phase separation of this mixture, and can hence inhibit the onset of phase separation altogether.
Further recent experiments have shown that when phase-separated domains are on the scale of network heterogeneities, they can take on a regular pattern structure with a well-defined length scale that does not coarsen over time.

In this manuscript, we have developed a poromechanical phase-field model that describes this phenomenon by considering a non-local elastic response to network deformation.
We have used linear stability analysis of our model to identify the parameters for which phase separation can occur in such a system, and to explore the conditions under which elasticity can inhibit phase separation.
Numerical simulations of our model reveal that non-local effects are key for arresting phase separation and inducing the formation of a patterned phase.
We find that phase separation initialises at small length scales determined by the most unstable mode of the system.
The resulting phase-separated mixture then partially coarsens to an equilibrium length scale that depends on the network stiffness.
In the absence of non-locality, the mixture instead coarsens completely towards a final state consisting of a single pair of phase-separated domains.

Our modelling approach utilises a large-deformation poromechanical framework that allows simulation of the dynamics of the process, building on the equilibrium approach previously developed by \citet{Qiang2024}.
Compared to the similar recent simulations of \citet{Oudich2025}, our model uses a fully non-linear description of the deformation kinematics, and considers non-local contributions to all components of the elastic energy.
In order to quantitatively compare the predictions of our model to experimental results, it will be essential to generalise the non-local elastic stress law described here to multiple spatial dimensions and more complex network rheologies~\citep{Fernandez-Rico2025}.
Nevertheless, the modelling effort and analysis presented here represents an important step towards understanding this physically and conceptually complex system.

\section*{Conflicts of interest}
There are no conflicts to declare.

\section*{Data availability}

MATLAB codes used to generate the results presented in this manuscript are available at Zenodo via \url{https://doi.org/10.5281/zenodo.16813489}. \nocite{Paulin2025Zenodo}

\section*{Acknowledgements}

The authors thank Filipe Thewes for helpful discussions about non-local interactions, and gratefully acknowledge funding from the Max Planck Society and the European Union (ERC, EmulSim, 101044662).

\appendix

\renewcommand{\theequation}{\thesection.\arabic{equation}}
\setcounter{equation}{0}

\section{Derivation of non-local stress law in differential form} \label{App:Stress}

For a Neo-Hookean--style non-local energy density of the form
\begin{equation} \label{app:PsiNL}
f\sub{nl} = \frac{G}{2} \left[ J \int J' g_\xi\intd{X}' - 1-2\textrm{ln}\tilde{J} \right] + \frac{K}{2} (J-1) \int (J' -1) g_\xi\intd{X'}  \;,
\end{equation}
the elastic stress is given in integral form as
\begin{equation} \label{app:SigmaNL}
\sigma = G \left[ \int J' g_{\xi}\intd{X'} - \frac{1}{\tilde{J}}\right] + K\int(J'-1)g_{\xi}\intd{X'} \; ,
\end{equation}
where $\tilde{J} =  \int J' g_\xi\intd{X}$.
For the convolution kernel $g_\xi=\exp({-|{X}|/\xi})/2\xi$, it is possible to reformulate \Eqref{app:SigmaNL} in a differential form.
First, we split the elastic stress into two parts such that $\sigma = \sigPrime - \sigStar$, where
\begin{equation} \label{app:SigPrime}
\sigPrime = \int_{-\infty}^\infty\left[ GJ' + K(J'-1) \right]g_\xi\left(|X'-X|\right)\intd{X'}
\end{equation}
and
\begin{equation} \label{app:SigStar}
\frac{1}{\sigStar} = \frac{1}{G} \int_{-\infty}^\infty J' g_\xi\left(|X'-X|\right)\intd{X} \;.
\end{equation}
Focussing first on $\sigPrime$, it is then convenient to split the integral in \Eqref{app:SigPrime} into two parts,
\begin{equation} \label{app:splitStress}
\sigPrime = \int_{-\infty}^X [GJ' + K(J'-1)] g_\xi^-\intd{X'} + \int_{X}^\infty  [GJ' + K(J'-1)] g_\xi^+\intd{X'}\;,
\end{equation}
where $g_\xi^- = \exp({-(X-X')/\xi})/2\xi$ and $g_\xi^+ = \exp({-(X'-X)/\xi})/2\xi$.
Differentiating \Eqref{app:splitStress} with respect to $X$ and using that $g_\xi^\pm(0) = 1/2\xi$, we then find
\begin{multline}
\pd{\sigPrime}{X} =   \int_{-\infty}^X  [GJ' + K(J'-1)]\pd{g_\xi^-}{X}\intd{X'} +  [GJ + K(J-1)] g_\xi^-(0)\\ 
 + \int_{X}^\infty  [GJ' + K(J'-1)]\pd{g_\xi^+}{X}\intd{X'} -  [GJ + K(J-1)]g_\xi^+(0) \; ,
 \end{multline}
 and hence
 \begin{equation}
\pd{\sigPrime}{X}  =  - \int_{-\infty}^X  [GJ' + K(J'-1)] \frac{g_\xi^-}{\xi}\intd{X'} + \int_{X}^\infty  [GJ' + K(J'-1)] \frac{g_\xi^+}{\xi}\intd{X'}\; .
\end{equation}
Similarly,
\begin{multline}
\pdd{\sigPrime}{X} = \int_{-\infty}^X [GJ' + K(J'-1)]\frac{g_\xi^-}{\xi^2}\intd{X'} - [GJ+ K(J-1)]\frac{g_\xi^-(0)}{\xi} \\+  \int_{X}^\infty [GJ' + K(J'-1)] \frac{g_\xi^+}{\xi^2}\intd{X'} - [GJ + K(J-1)]\frac{g_\xi^+(0)}{\xi}  \;
\end{multline}
and therefore
\begin{equation}
\pdd{\sigPrime}{X} = \frac{1}{\xi^2} \left[ \int_{-\infty}^\infty [GJ' + K(J'-1)] g_\xi \intd{X'} - [GJ + K(J-1)] \right] \; .
\end{equation}
Inserting the definition of $\sigPrime$ and rearranging, we thus find
\begin{equation} \label{app:sigPrimeEq}
\left(1- \xi^2\pdd{}{X}\right)\sigPrime = GJ + K(J-1) \; .
\end{equation}
Repeating this procedure for \Eqref{app:SigStar} correspondingly yields
\begin{equation} \label{app:sigStarEq}
\left(1- \xi^2\pdd{}{X}\right)\frac{1}{\sigStar} = \frac{J}{G} \;.
\end{equation}
Given $\sigPrime$ and $\sigStar$ from \Eqsref{app:sigPrimeEq} and \eqref{app:sigStarEq}, the total elastic stress is then ${\sigma = \sigPrime - \sigStar}$.

\bibliographystyle{plainnat}
\bibliography{nonlocal_dynamics.bib}

\end{document}